\documentclass[printer]{aa}

\newcommand{\out}[1]{}  
\newcommand{\onine}{\hbox{0943--242}} 
\newcommand{\otwo}{\hbox{0200+015}} 
\newcommand{\etal}{\hbox{et~al.}}
\newcommand{\msol}{\hbox{${\rm M_{\sun}}$}}

\newcommand{\qtscm}{\hbox{${\rm quanta}\,{\rm cm^{-2}}\,s^{-1}$}}
\newcommand{\qts}{\hbox{${\rm quanta}\,s^{-1}$}}
\newcommand{\cms}{\hbox{${\rm cm^{-2}}$}}

\newcommand{\cmc}{\hbox{${\rm cm^{-3}}$}}
\newcommand{\kms}{\hbox{${\rm km\,s^{-1}}$}}
\newcommand{\inu}{\hbox{${\rm erg}\,{\rm cm}^{-2}\,{\rm s}^{-1}\,{\rm Hz}^{-1}\,{\rm sr}^{-1}$}}

\newcommand{\fla}{\hbox{${\rm erg}\,{\rm cm}^{-2}\,{\rm s}^{-1}$\,\AA$^{-1}$}}
\newcommand{\ergs}{\hbox{${\rm erg\, s^{-1}}$}}

\newcommand{\mjy}{\hbox{$\mu$Jy}}
\newcommand{\up}{\hbox{\it U}}
\newcommand{\upvv}{\hbox{$U_{0.1}$}}
\newcommand{\upn}{\hbox{$U_{0.1}^{\rm thin}$}}
\newcommand{\upk}{\hbox{$U_{0.005}^{\rm thick}$}}
\newcommand{\etap}{\hbox{$\eta$}}
\newcommand{\fco}{\hbox{F$^{\rm obs}_{\rm c}$}}
\newcommand{\flyo}{\hbox{F$^{\rm obs}_{\rm Ly\alpha}$}}

\newcommand{\ewr}{\hbox{EW$^{\rm rest}_{\rm Ly\alpha}$}}

\newcommand{\jnu}{\hbox{$J_{\nu}$}}
\newcommand{\jto}{\hbox{$J_{-21}$}}
\newcommand{\teff}{\hbox{$T_{\rm eff}$}}
\newcommand{\nheii}{\hbox{$N_{\rm HeII}$}}
\newcommand{\at}{\hbox{$A_{0.2}$}}
\newcommand{\rt}{\hbox{$r_{25}$}}
\newcommand{\rn}{\hbox{$r^{\rm thin}_{\rm kpc}$}}
\newcommand{\rk}{\hbox{$r^{\rm thick}_{\rm kpc}$}}
\newcommand{\nvv}{\hbox{$n_{0.01}$}}
\newcommand{\nn}{\hbox{$n_{\rm H}^{\rm thin}$}}
\newcommand{\nk}{\hbox{$n_{\rm H}^{\rm thick}$}}
\newcommand{\phih}{\hbox{$\varphi_{\rm H}$}}
\newcommand{\qh}{\hbox{$Q_{\rm H}$}}

\newcommand{\nnh}{\hbox{$n_{\rm H}$}}
\newcommand{\nh}{\hbox{$N_{\rm H}$}}
\newcommand{\mh}{\hbox{$M_{\rm H}$}}
\newcommand{\mhn}{\hbox{$M_{\rm H}^{\rm thin}$}}
\newcommand{\mhk}{\hbox{$M_{\rm H}^{\rm thick}$}}
\newcommand{\nhn}{\hbox{$N_{\rm H}^{\rm thin}$}}
\newcommand{\nhk}{\hbox{$N_{\rm H}^{\rm thick}$}}
\newcommand{\nhi}{\hbox{$N_{\rm HI}$}}

\newcommand{\nciv}{\hbox{$N_{\rm CIV}$}}
\newcommand{\nchr}{\hbox{$N_{\rm CIV}/N_{\rm HI}$}}
\newcommand{\nchrl}{\hbox{$N_{\rm CII}/N_{\rm HI}$}}
\newcommand{\nnhr}{\hbox{$N_{\rm NV}/N_{\rm HI}$}}
\newcommand{\nohr}{\hbox{$N_{\rm OVI}/N_{\rm HI}$}}
\newcommand{\nmghr}{\hbox{$N_{\rm MgII}/N_{\rm HI}$}}
\newcommand{\map}{\hbox{{\sc mappings i}c}}
\newcommand{\lan}{\hbox{{\sc lan}}}
\newcommand{\mbr}{\hbox{{\sc mbr}}}
\newcommand{\sed}{\hbox{{\sc sed}}}
\newcommand{\seds}{\hbox{{\sc sed}{\rm s}}}
\newcommand{\eelr}{\hbox{{\sc eelr}}}
\newcommand{\eelrs}{\hbox{{\sc eelr}{\rm s}}}
\newcommand{\hzrg}{\hbox{{\sc h}\hspace{-1.2pt}{\it z}{\sc rg}}}
\newcommand{\hzrgs}{\hbox{{\sc h}\hspace{-1.2pt}{\it z}{\sc rg}{\rm s}}}

\newcommand{\voutp}{\hbox{$V_o^{\prime}$}}

\newcommand{\epsnagn}{\hbox{${\epsilon}_{\rm neb}^{\rm AGN}$}}
\newcommand{\epsn}{\hbox{${\epsilon}_{\rm neb}$}}
\newcommand{\epsnt}{\hbox{${\epsilon}_{0.5}^{\rm neb}$}}

\newcommand{\epsh}{\hbox{${\epsilon}_{\rm shell}$}}

\newcommand{\ze}{\hbox{$z_{\rm e}$}}
\newcommand{\za}{\hbox{$z_{\rm a}$}}
\newcommand{\zsol}{\hbox{$Z_{\sun}$}}
\newcommand{\zstar}{\hbox{$Z_{\ast}$}}

\newcommand{\mgiiw}{\hbox{Mg\,{\sc ii}\,$\lambda\lambda $2798}}
\newcommand{\mgii}{\hbox{Mg\,{\sc ii}}}

\newcommand{\ciiw}{\hbox{C\,{\sc ii}$\lambda\lambda $1335}}
\newcommand{\cii}{\hbox{C\,{\sc ii}}}

\newcommand{\civ}{\hbox{C\,{\sc iv}}}
\newcommand{\civw}{\hbox{C\,{\sc iv}\,$\lambda\lambda $1549}}
\newcommand{\civww}{\hbox{C\,{\sc iv}\,$\lambda\lambda $1548, 1551}}

\newcommand{\nivw}{\hbox{N\,{\sc iv}]$\,\lambda $1485}}
\newcommand{\nv}{\hbox{N\,{\sc v}}}
\newcommand{\nvw}{\hbox{N\,{\sc v}$\lambda $1240}}
\newcommand{\ovi}{\hbox{O\,{\sc vi}}}
\newcommand{\oviw}{\hbox{O\,{\sc vi}$\lambda $1035}}

\newcommand{\osiivw}{\hbox{O\,{\sc iv}+Si\,{\sc iv}$\lambda $1402}}

\newcommand{\Lya}{\hbox{$L_{\rm Ly\alpha}$}}

\newcommand{\lya}{\hbox{Ly$\alpha$}}

\newcommand{\hi}{\hbox{H\,{\sc i}}}
\newcommand{\hii}{\hbox{H\,{\sc ii}}}

\newcommand{\heii}{\hbox{He\,{\sc ii}}}

\newcommand{\heiiw}{\hbox{He\,{\sc ii}\,$\lambda $4686}}
\newcommand{\heiiuw}{\hbox{He\,{\sc ii}\,$\lambda $1640}}

\usepackage{graphics} 

\begin{document}



\title{Ionization of large-scale absorbing haloes and feedback events from
high-redshift radio galaxies}


\author{L. Binette\inst{1}, R. J. Wilman\inst{2}, 
M. Villar-Mart\'\i n\inst{3},  
R.A.E. Fosbury\inst{4} 
M. J. Jarvis\inst{5}
and H. J. A. R\"ottgering\inst{6} 
          }

\authorrunning{Binette et~al.}

\offprints{R.J. Wilman}

\institute{Instituto de Astronom\'\i a, UNAM,  Ap. 70-264, 04510 M\'exico, 
DF, M\'exico 
\and Department of Physics, University of Durham, South Road, Durham DH1 3LE, UK~(e-mail: r.j.wilman@durham.ac.uk)  
\and Instituto de Astrof\'\i sica de Andaluc\'\i a, CSIC, Apdo. 3004, 18080, Granada, Spain
\and ST-ECF, Karl-Schwarzschild Strasse 2, D-85748 Garching bei M\"unchen, Germany
\and Astrophysics Department, Keble Road, Oxford OX1 3RH, UK
\and Leiden Observatory, P. O. Box 9513, 2300 RA Leiden, The Netherlands
}
\date{Received February 2006/ Accepted}

\titlerunning{\hzrg\ haloes}

\authorrunning{Binette  et al.}

\abstract{}{We present photoionization calculations for the spatially
extended absorbers observed in front of the extended emission line
spectrum of two high redshift radio-galaxies: \onine\ ($\ze=2.922$)
and \otwo\ ($\ze=2.230$), with the aim of reproducing the absorber
column ratio, \nchr.}{We explore the effects of using different UV
continua in the photoionization calculations. A comparison is drawn
between the absorber in \otwo\ and the two absorbers observed near the
lensed Lynx Arc Nebula at redshift 3.36, which present very similar
\nchr\ ratios.}{We find that hot stars from a powerful starburst, or a 
metagalactic background radiation (\mbr) in which stars dominate over quasars, 
are equally successful in
reproducing the observed \nchr, assuming subsolar gas metallicities
for each absorber. These softer \seds\ eliminate the difference of a
factor 1000 in metallicity between the two absorbers encountered in
earlier work where a powerlaw \sed\ was assumed. 
The detection of continuum flux in \onine\ suggests that the level of
ionizing photons is consistent with a stellar ionizing source. }
{If the \mbr\ is responsible for the ionization of
the radiogalaxy absorbing shells, their radii (if spherical) would be
large ($> 100\,$kpc) and their mass would be huge $>10^{12}\,$\msol,
implying that the feedback mechanism initiated by the central galaxy
has caused the expulsion of more baryonic mass than that left in the
radiogalaxy.  If, as we believe is more likely, stellar ionizing sources within
the radio galaxy are responsible for the absorber's ionization,
smaller radii of $\sim 25$~kpc and much smaller masses ($\sim 10^{8} -
10^{10}$\msol) are inferred. This radius is consistent with 
the observed transition in radio source size between the smaller sources in which 
strong \hi\ absorption is almost ubiquitous, and the larger sources where it is
mostly lacking. 
Finally, we outline further absorption line diagnostics which could be used to 
constrain further the properties of the haloes and their source of ionization.}

\maketitle

\section{Introduction}\label{sec:intr}

A prominent characteristic of high-redshift radio-galaxies (\hzrgs) at
$z> 2$ is their spatially extended line emission regions (hereafter
\eelr), which are often luminous in \lya\ ($>10^{44}\,$\ergs) and
extended over several to tens of kpc. The excitation mechanism for the
{\it emission gas} is either shock excitation by jet material or AGN
photoionization (the presence of \nvw\ line emission precludes stellar
photoionization). The \eelr\ is kinematically active, with FWHM
reaching $1000\,$\kms.  With observations of a sample of \hzrgs,
Van\,Ojik et\,al. (1997;VO97) discovered that, when observed at
intermediate resolution (1--2\,\AA), the majority of \hzrgs\ with
small radio source sizes ($<50$\,kpc) exhibit narrow \lya~\hi\
absorption. This absorption is superimposed upon the \lya\ emission
with a spatial extent comparable to that of the \eelr. In addition to
\lya, the \civw\ doublet has also been observed in absorption in two
\hzrgs, superimposed on the \civ\ emission line, firstly in \onine\
($\ze=2.922$) (Binette et\,al. 2000, hereafter B00) and secondly in
\otwo\ ($\ze=2.230$) (Jarvis et\,al. 2003, hereafter J03). Building on
the results of B00 and J03, in the present paper we examine the
excitation mechanism of the large scale absorbing haloes in greater
detail by exploring photoionization by a variety of different spectral
energy distributions (hereafter \sed).

The basic structure of the paper is as follows. In the remainder of section 1 we 
review our current understanding of \hzrg\ absorbers, focussing on the distribution, ionization and metallicity of the absorbing gas and the specific problems which motivate our current study; an insightful comparison is made with the absorbers in the Lynx Arc Nebula (\lan), a gravitationally-lensed \hii\ galaxy at $z=3.357$. In section 2 we summarise the observational results we aim to reproduce, namely the \nchr\ ratio in the aforementioned \hzrgs\ and the \lan. Section 3 describes the \map\ code and our assumptions concerning the photoionizing \seds. Section 4 presents the results of these calculations and in section 5 we assess their implications for the origins of the absorbers and their compatibility with other observables. Finally, in section 6 we present some additional absorption line diagnostics which may in future help to discrimate between the proposed scenarios.

\subsection{Shell-like structure for the \hzrg\ haloes} \label{sec:shel}

Among the \hzrgs\ with small radio sources ($<50$\,kpc), the detection
rate of associated absorption systems is 90\% (9 out of 10 \hzrgs\ in the V097 
study) while it is  only 25\% for  larger
radio-sizes. The fact that the absorption extends over the whole background
\eelr\ emission favours a shell-like geometry for the absorption systems
rather than a conglomerate of individual clouds, as proposed initially
by VO97.  In Sect.\,\ref{sec:geo} we give further indications as to
why we retain the simplifying assumption of a simple shell structure
in the current work. Because the density per unit redshift of the
strong absorbers ($\nhi > 10^{18}\,$\cms) around \hzrgs\ was found to be
much higher than that given by the statistics of intergalactic medium
(IGM) absorbers at large, VO97 inferred that they belong to the
environment of the parent \hzrg\ rather than to the IGM. The density
of the thinnest absorbers ($< 10^{15}\,$\cms) around \hzrgs\ on the
other hand is comparable to that of \lya\ forest absorbers in the IGM,
as more recently shown by Wilman et\,al. (2004: W04).  It is
conceivable that the physical conditions in the
thin\footnote{By ``thin'' we refer to a small \hi\ {\it column density}, not to a small physical size.}
\hzrg\ absorbers are indistinguishable from those operating within
typical IGM \lya\ forest absorbers. The available data, however, are
still insufficient to confirm or refute this proposition.

The rarity of absorbers among \hzrgs\ with radio-sizes
larger\footnote{VO97 (p. 369) showed that this finding is not the
result of a selection effect due to the fact that  larger radio sources
have a narrower \lya\ emission profile.} than 50\,kpc suggests
that the typical lateral dimensions of the shell (in the plane of the sky)
might be $\la 50$\,kpc. The proposed interpretation is that, as the
AGN jet expands beyond such size, the bow-shocks overtake the shells
and disrupt them. This is the first scenario, which we label A or
``inner shell scenario''. If valid, it suggests that the expansion of
the AGN {\it jet cocoon} is not the mechanism by which the shells are
formed, but rather by which they are destroyed.  Scenario A favours a
shell formation mechanism that relies on large-scale outflows
generated by episodes of massive star formation.  Using high
dispersion data from VLT-UVES, W04 proposed that the absorbers in
\hzrgs\ probably lie within the core of young galactic protoclusters,
consistent with observations of their environments (e.g. Venemans et
al. 2005; Overzier et al. 2006),
and may be a byproduct of massive galaxy formation. Krause (2005)
published hydrodynamical simulations of the formation of a shell due
to the expansion of a stellar wind bow-shock. At a later stage in his
model, an AGN jet is launched and a jet cocoon builds up.  Once the
jet has extended beyond the initial bow-shock, the jet cocoon destroys
the shell as it overtakes it.
An estimate of the timescale for this to occur can be obtained if one follows the reasoning of J03,  where the radio-size represents
a kind of internal clock (see Sect.\,\ref{sec:bck}), which
characterizes not only the radio jet's age  but
also that of the starburst superwind that generates the shells.

A second possibility is that the rarity of shells among \hzrgs\ with
large radio-sizes may reflect an older phase in which the shells
have expanded further out and thinned out considerably. This process
would eventually render them undetectable (using the VO97 detection
technique) when their \nhi\ columns drops below $\la
10^{13}\,$\cms. This is the second scenario, which we label B or
``aging shell scenario''. In this case, the distance between the shell
and the parent \hzrg\ is unknown and can be much larger than the upper
limit size implied by scenario A, as will be discussed in
\ref{sec:sizm}. Scenario B leaves the
possibility open that some of the shells may result from the expansion
of a jet bow-shock (e.g. Krause 2002), although the most likely
formation mechanism of the shells remains stellar superwind, as in
scenario A.

The large scale \hzrg\ absorbers might be analogous to the absorbers detected
within\footnote{Where $h=H_o/100$\,km\,s$^{-1}$\,Mpc$^{-1}$.}
20--50$h^{-1}$\,kpc of high redshift galaxies, by Adelberger
et\,al. (2005) using nearby-field spectroscopy of background QSOs or
galaxies. The advantage of \hzrg\ absorber studies is that the
intrinsic shell outflow velocity is more readily available from
observations, but not their distance from the parent \hzrg\ (the
reverse applies to the technique used by Adelberger et\,al., assuming
spherical expansion of the absorbers).

The \hzrg\ shells share many similarities with at least some of
\lya-emitting `blobs' (hereafter LAB; Steidel et\,al. 2000) 
associated with Lyman Break Galaxies (Pettini et\,al. 2001), and
which are characterized by an \eelr\ that can reach large sizes
of up to $\sim 100$\,kpc. An important difference is that the radio
luminosities  are much fainter or undetected in the latter case. Using integral
field spectroscopy, Wilman et\, al. (2005) observed the \lya-emitting `blob' LAB-2 
in the SSA22 protocluster at $\ze=3.09$; they discovered a 
foreground absorber ($\nhi \simeq 10^{19}\,$\cms) with remarkable velocity coherence over a projected size of $\sim 76 \times 26$\,kpc.  Their interpretation is that a {\it galaxy-wide} superwind has swept up $\sim 10^{11}\,$\msol\ of diffuse material from the IGM over a few $10^{8}$~yr. This is a manifestation of the `feedback' mechanism thought to be regulating the formation of galaxies.



\subsection{What is ionizing the  large scale haloes?} \label{sec:ion}

At least a fraction of the known haloes appear to be highly ionized. 
In both \hzrgs\ in which a \civw\ doublet has been observed in
absorption (\onine\ and \otwo), the absorption redshift corresponds to 
one of the \lya\ absorbers.  
These two absorbing haloes therefore contain ionization species up to
at least C$^{+3}$.  B00 and J03 assumed that the \hi\ and
\civ\ absorption species occur within a physically contiguous
 structure, an aspect discussed further in Sect.\,\ref{sec:geo}. 

The possibility that the \hzrg\ absorbing haloes are photoionized by
the hidden nuclear radiation can be ruled out. Firstly, because there no 
observed continuum of sufficient strength underlying the \eelr. This 
is as expected in the quasar-radio-galaxy unification 
picture (Barthel 1989; Antonucci 1993; Haas et\,al. 2005), in
which the nuclear ionizing radiation is collimated along two
ionization cones, which in radio-galaxies lie along the plane of the
sky, and is therefore invisible to the observer and presumably to the
intervening absorbers as well, unless rather contrived gas geometries
are postulated. Secondly, B00 showed that the \civ/\lya\ emission- and
absorption-line ratios in \onine\ could not be reconciled with any
model in which the absorption- and emission gas are co-spatial. They
concluded that the absorbing gas is of much lower metallicity and
located further away from the host galaxy than the \eelr. Although
they favoured the idea that the diffuse {\it metagalactic  background
radiation} (hereafter \mbr) rather than the parent AGN was responsible 
for ionizing the absorbing haloes, in their calculations B00 and J03 
used a simple powerlaw as a crude approximation of the \mbr\ energy 
distribution. In this paper, we will assume more realistic \seds\ that 
take into account the cumulative opacity of IGM Lyman limit systems and \lya\ forest absorbers.

As for the possibility of collisional ionization of the shells, we
indicated in J03 that this mechanism was unlikely in the case of
\otwo, and that steady-state photoionizing shocks (Dopita \&
Sutherland 1996) resulted in rather large \nhi\ columns ($\sim 10^{19}\,$\cms),
incompatible with the small value characterizing \otwo. In the case
of \onine, the near solar metallicity models of Dopita
\& Sutherland (1996) do not attain the observed \nciv\ value 
for shock velocities below 400\,\kms\ and above this velocity the
\nhi\ column becomes excessive, requiring   the shock structure  to
be truncated. As for the photoionized precursor nebula upstream from
the shocks, the \sed\ generated downstream by fast shocks is as hard
as a powerlaw of index $\alpha \simeq -0.5$ up to $\ga 500$\,eV
(Binette, Dopita \& Tuohy 1985). Therefore, photoionization calculations
with a powerlaw as  presented in Sect.\,\ref{sec:agn} capture
the main features of such a precursor. In essence,  any hard \sed\
requires supersolar metallicities in order to fit the column ratios found in
\otwo. Finally, calculations to represent the case of a collisionally 
ionized gas slab at temperature $T$ has been explored by J03. They
find that for \otwo, the \nchr\ column ratio could be reproduced by
using roughly solar metallicities, provided that $T$ is finetuned to lie
around $10^5$\,K.  Apart from the fact that this metallicity is rather
large for the redshift considered, it would be difficult to explain
how the plasma could be maintained at a temperature approaching the
peak of its cooling curve. This  would most likely require a yet unknown
heating mechanism.

\subsection{The  case for  a simple scattering screen }  \label{sec:geo}

In a morphologically and kinematically complex \eelr, one cannot readily 
disentangle photon destruction due to line-of-sight absorption from the effects
of transmission by multiple scatterings. Nevertheless, for the large scale 
absorbing haloes in \otwo\ and \onine\ (or other \hzrgs\ 
studied by VO97), there is no evidence that the absorbers share the complexity 
of the \eelr. The results suggest that a uniform foreground scattering screen 
provides an adequate description of the ``absorbing'' haloes in \hzrgs. 
Strong evidence for this was provided by observations at much higher spectral 
resolution using VLT-UVES (e.g. J03 and W04). In particular, J03 finds that the main 
absorber in \onine\ remains as a single system of column density $\sim
10^{19}\,$\cms\ over the full size of the \eelr, being completely
black at its base, with no evidence for substructure or a multiphase
environment. The absorption trough is blueshifted by 265\,\kms\ 
with respect to the centroid of
the background emission profile. This spatial and kinematical
coherence of the absorber contrasts with the chaotic multiphase medium 
encountered in the Galactic ISM or the \eelr\ of \hzrgs. It also suggests that
the absorber is physically separate from the background \eelr\ and that it is 
therefore simply acting as a scattering surface, as argued in J03. This clean 
separation between \eelr\ and the absorber simplifies the modelling task and 
justifies the ionization-stratified slab approximation adopted in Sect.\,\ref{sec:resu}.

\subsection{Comparison with the Lynx Arc Nebula: \lan} \label{sec:lan}

Following an independent study of the lensed Lynx Arc
Nebula{\footnote{The Lynx Arc Nebula is a high-redshift metal-poor
gravitationally lensed \hii\ galaxy that was discovered
serendipitously by Holden et\,al.  (2001).} at
$z=3.357$ by some of us (Fosbury et al. 2003), it was observed that the 
column ratios \nchr\ in the \lan\ and \otwo\ are very similar.
In the calculations which follow, we therefore explicitly compare the \lan\ and 
the \hzrgs\ absorbers, making use of the following insights which place the physical 
conditions in the \lan\ on a firm footing. Firstly, the \lan\ is an active star-forming object, so we may reasonably assume that the subsolar metallicity that
characterizes the emission gas, $\sim 10$\% (following the work of VM04), also 
applies to the absorbing gas. Secondly, the \lan\ presents a relatively high 
excitation {\it emission line} UV spectrum, which photoionization by hot stars can
reproduce successfully.  Photoionization by a straight
powerlaw{\footnote{The possibility that the \lan\ corresponded to
photoionization by a quasar powerlaw that is partly absorbed has been
investigated by Binette et\,al. (2003: BG03). In order to work, such
an absorbed powerlaw model requires a very specific and unlikely fine
tuning of the parameters defining the
\eelr\ and the hidden filtering screen. }}, on the other hand, would
result in the emission of a detectable \nvw\ line (comparable in strength to  
\nivw\ line, see BG03), which is not observed. Hence, for this object, we know the 
absorber metallicity and excitation source with some
confidence. Therefore, the successful reproduction of the \lan\ column
ratios in Sect.\,\ref{sec:stel} using subsolar metallicities and
photoionization by hot stars, prompts us to consider that such an
\sed\ might also apply to the \hzrg\ absorbers.

\subsection{Metallicity evolution vs. softer  ionizing \sed} \label{sec:bck}

With the VLT-UVES, J03 obtained superb spectra of the aforementioned \hzrgs\ at ten 
times the resolution used by VO97. The spectra confirmed that the main
absorber in \onine\ exhibits no additional substructure to that
reported by VO97, as already discussed. In contrast, a very different view of \otwo\
emerges: the single absorber with HI column density $\sim
10^{19}\,$\cms\ seen at low resolution now splits into two $\sim 
10^{14.6}\,$\cms\  systems; these extend by more than 15\,kpc to obscure
additional \lya\ emission coincident with a radio lobe. Additional but fragmented
 absorbers are seen on the red wing of the emission
line at this position.  We recall that gas metallicities as high as
$\sim 10\,$\zsol\ are required to reproduce the \nchr\ ratio in
\otwo\ (Sect.\,\ref{sec:bck}; J03) assuming photoionization by a
straight powerlaw. This suggests that the absorbing gas has
undergone very substantial metal enrichment.  Based on the smaller
radio source size in \onine\ (26\,kpc versus 43\,kpc for \otwo), 
J03 conjectured that the radio source age (as inferred from
its linear size) is the parameter controlling the evolution of:
(i) the structure/kinematics of the absorbing halo, through
interaction and shredding of the initially quiescent shells; and (ii)
its metallicity built-up, through enrichment by starburst superwind
triggered concurrently with the nuclear radio source.

Although this age and enrichment scenario (B) remains an appealing
possibility, the large metallicity gap inferred by J03 of three
orders of magnitude between \onine\ ($\sim 0.01\,$\zsol) and
\otwo\ ($\sim 10\,$\zsol) is a cause for concern. Here we revisit the issue by
exploring alternative ionizing \seds\ that would require only a factor
of ten metallicity enhancement with respect to 0943$-$242. We focus on the case of 
\seds\ from hot stars and the diffuse \mbr\ with the aim of generating a grid of 
models for comparision with future observations. 




\section{The observational dataset}\label{sec:data}

In this section, we gather together the principal observational results 
which we aim to reproduce, namely the \hi\ and \civ\ column densities for 
the absorbers in the two \hzrgs\ and the \lan.

The \civ\ and the \lya\ absorption columns in \onine\ ($\ze=2.922$)
and \otwo\ ($\ze=2.230$) have been measured by various authors
(R\"ottgering et al. 1995; B00, J03, W04). We adopt the values of J03,
which are based on VLT-UVES observations of both \hzrgs.  In
\onine\ the dominant large scale absorber is characterized by an
\nhi\ column of $10^{19.1}\,$\cms, which puts it among the group of
larger \hi\ columns (see W04). However, the four \lya\ absorbers
observed in \otwo\ are rather thin, with columns of order
$10^{14.7}$\,\cms. These all belong to the group of smaller
\hi\ columns haloes, which are much more numerous  (see W04). 
In \onine, the \civww\ doublet is observed in absorption at the
same redshift as the dominant \hi\ absorber (J03; B00; R\"ottgering \&
Miley 1997) and corresponds to a column of $10^{14.6}$\,\cms. In the
case of \otwo, only one \hi\ absorber, with $\nhi =
10^{14.7}$\,\cms\ shows a corresponding \civ\ doublet in absorption,
with $\nciv=10^{14.6}$\,\cms.  The \nchr\ column ratios for
\onine\ and \otwo\ are $10^{-4.5}$ and $10^{-0.07}$, respectively.

As for the \lan, the two local absorption systems in the Lynx Arc have
been labelled {\it a1} and {\it a2} by Fosbury et\,al. (2003) who
determined the \hi\ column to be $1.05 \times 10^{15}$ and $0.60
\times 10^{15}$\,\cms, respectively, and the \civ\ columns to be $0.83
\times 10^{15}$ and $1.02 \times 10^{15}$\,\cms, respectively.  The
\nchr\ column ratios for {\it a1} and {\it a2} are therefore
$10^{-0.10}$ and $10^{0.23}$, respectively. The similarity of {\it a1}
with \otwo\ is noteworthy.

\section{Photoionization models and ionizing energy distributions}\label{sec:sed}

\subsection{\map\ and gas abundances}\label{sec:abu}

To compute the \nchr\ ratio, we have used the code \map\ (Binette,
Dopita \& Tuohy 1985; Ferruit et~al 1997).  To represent  solar
 abundances, we adopt the set of Anders \& Grevesse (1989).  When varying 
metallicities, we multiply the solar abundances of all elements heavier than He
by a constant, which we label the gas metallicity (in units of
\zsol). For the \hzrg\ absorber, we assume a slab geometry illuminated on one side. 
For each ionizing \sed\ that we considered, we calculated the
equilibrium ionization state of the gas and integrated the ionization
structure inward until a preselected target value of the column \nhi\
was reached. For the range of parameters explored in this paper -- where the aim is
to reproduce the observed \nhi\ and the \nchr\
column ratios -- all models of the absorbers turn out to be matter-bounded (\otwo) or
marginally optically thick to the ionizing radiation (\onine).

\subsection{Continuum softness}\label{sec:soft}

For a given input \sed, the calculations are repeated for different
values of the ionization parameter\footnote{We use the customary
definition of the ionization parameter $\up\ = \phih/c \nnh$ as the
ratio between the density of ionizing photons impinging on the slab
$\phih/c$ and the total H density at the face of the slab \nnh.}
in order to build a sequence of models in \up,  starting at the minimum
value of $0.001$. 

It is customary to define the \sed's softness using the parameter
\etap, which is  the column ratio of singly ionized He to neutral H, \nheii/\nhi. This ratio
does not, however, uniquely define the \sed, as \etap\ also depends on the
slab thickness and on \up, and not just on the continuum's shape (see
for instance Appendix\,A of Fardal et\,al. 1998).  In the case of
stellar \seds, \etap\ varies rather abruptly with \teff.  For
instance, $\etap$ is 1480 for a 71\,000\,K star while for a 80\,000\,K
star its value is only 95.  Next to each \sed\ in Fig.\,\ref{fig:sed},
we indicate   the value of \etap\ calculated between brackets, assuming
$\nhi\ = 10^{14.8}\,$\cms\ and $\up=0.1$. 
We now review the various \seds\ displayed in Fig.\,\ref{fig:sed} and
used in the calculations reported in Sec.\,\ref{sec:resu}.

\begin{figure}    
\resizebox{!}{14cm}{\includegraphics{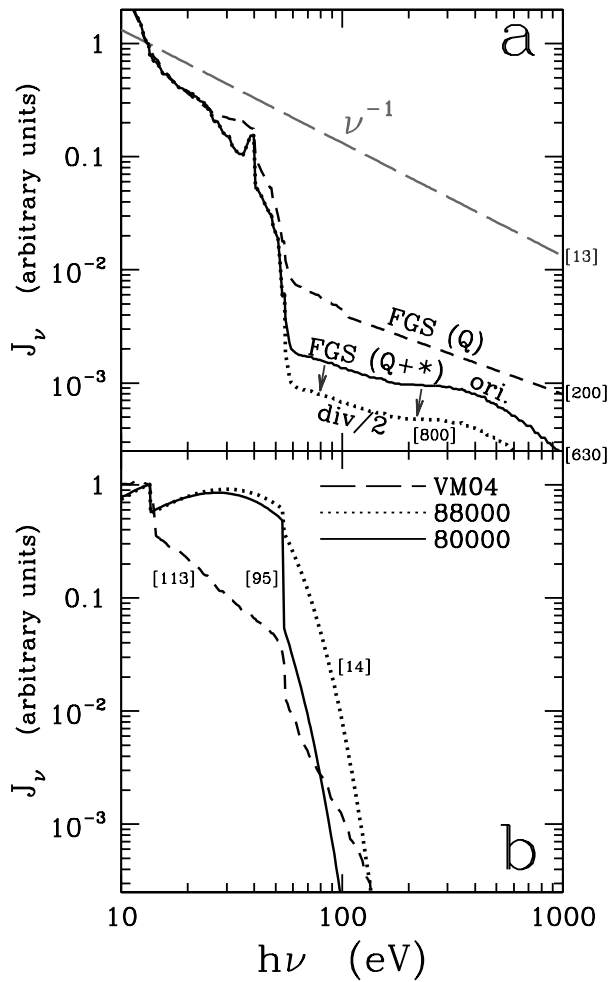}}
\caption{The spectral energy distribution of various 
ionizing sources (see Sect.\,\ref{sec:sed}) as a function of photon
energy.  Panel a: silver long-dashed line: spectral energy
distribution corresponding to an AGN powerlaw, short dashed-line:
diffuse \mbr\  energy distribution from FGS at $\za=2$ comprising only quasars (Q) as
sources, continuous line: (original) \mbr\ energy distribution  from FGS at $\za=3$
comprising both quasar and stellar sources (Q+{\Large\mbox{$\star$}}), dotted line:
same as solid line except that the flux beyond 54.4\,eV has been
reduced by a factor two. Panel b: continuous line: \sed\ of a
zero-age metal-free star of effective temperature 80\,000\,K,
dotted-line: \sed\ of a zero-age metal-free star of effective
temperature 88\,000\,K, long-dashed line:
\sed\ from an evolutionary model of CMK04
corresponding to an age of 3.4\,Myr and metallicity of 20\% solar. 
Values of \etap\ are given between brackets for each \sed\ (Sec.\,\ref{sec:soft}).}
\label{fig:sed}
\end{figure} 

\subsection{AGN powerlaw \sed}\label{sec:pow}

In the case of direct photoionization by an AGN,  we assume a
simple power law of index $\alpha = -1.0$ as in B00  (with $\jnu \propto
\nu^{+\alpha}$). This \sed\ is represented by the gray long-dashed line
in Fig.\,\ref{fig:sed}a.

\subsection{\sed\ of the diffuse metagalactic  radiation  (\mbr)}\label{sec:meta}

The integrated ultraviolet flux arising from distributed QSOs and/or
from hot massive stars (in metal-producing young galaxies) is believed to
be responsible for maintaining the intergalactic diffuse gas, the
\lya\ forest and Lyman limit systems in a highly ionized state. The
spectrum and intensity of the diffuse \mbr\ is affected not only by
photoelectric absorption from intergalactic matter, but also by the
re-emission from radiative recombinations within the absorbing gas
itself.  In short, QSO absorption-line systems are sources, not just
sinks of ionizing photons, as shown by Haardt \& Madau (1996).

Detailed calculations of the propagation of  QSO and
stellar ionizing radiation through the intergalactic space have been
presented by Fardal et\,al. (1998; FGS in the figures or footnotes) and the 
resultant \seds\ relevant to the current work\footnote{The \seds\ calculated by FGS are 
 softer than those of Haardt \& Madau (1996). In their
Appendix\,A, FGS justifies this difference on account of a more
detailed treatment of the cloud opacity and re-emission.}  are shown in
Fig.\,\ref{fig:sed}a.  On the one hand, we have the metagalactic \sed\
in which {\it only} quasars are contributing (short dashed line) corresponding 
to  model Q2 in their Fig.\,7 and, on the other, the
\sed\  in which hot stars from star forming regions
are included (continuous line), a
model shown in their Fig.\,6. In this model, the stars are
contributing twice the flux of quasars at 13.6\,eV. The dotted line is
a similar \sed, except that the flux beyond 4\,Ry has been divided by
two.  It is an ad\,hoc model representing the case in which stars are
contributing proportionally more with respect to quasars (a similar
\sed\ was also considered by Telfer et\,al. 2002).


The sharp drop of flux at 54.4\,eV is a characteristic of all
metagalactic radiation models and is due to the cumulative opacity of \heii\
within the IGM.  As the IGM \seds\ extend into the soft X-rays, it is
important to include the harder radiation beyond $\ga 200$\,eV, otherwise
the calculated \civ\ columns are affected, especially when
\up\ is large.

If we turn to the values of the softness parameter
(Sect.\,\ref{sec:soft}) observed among IGM absorbers, there is a
substantial dispersion in the values measured by Kriss et\,al. (2001),
with $1 \la \etap \la 1000$, which suggests that for a fraction of
absorbers, stellar ionizing sources might be contributing. The
possibility of a rather inhomogeneous distribution of the \sed\ 
hardness according to location is favoured by the independent study of
Smette et\,al. (2002), who find that $20 \la \etap \la 5000$.

\subsection{Stellar ionizing \seds}\label{sec:str}

In the stellar ionizing case (panel b in in Fig.\,\ref{fig:sed}), we
considered metal-free stellar \seds\ that approximate those studied by Schaerer
(2002) with \teff\ among one of the following
values: 42\,000, 57\,000, 71\,000, 80\,000 and 88\,000\,K.  In
Fig.\,\ref{fig:sed}b, we illustrate the cases of the 80\,000 and
88\,000\,K \seds\  (the continuous and dotted lines, respectively).
As in BG03,  who presented various photoionization models for the \lan, we
approximate the selected stellar \seds, using a technique that
reproduces the ionizing photon luminosities $Q({\rm H})$, $Q({\rm
He^0})$ and $Q({\rm He^+})$ of the selected ${\rm log}\,T_{\rm eff}$
model listed in Table\,3 of Schaerer (2002). In a similar fashion to
Shields \& Searle (1978), we derive the monochromatic temperatures at
the edge boundaries $T_{\rm H^0}^+$, $T_{\rm He^{+}}^-$ and $T_{\rm
He^{+}}^+$, and then interpolate linearly in ${\rm log} \; T_{\nu}$
for all the wavelengths used in the code \map. We equated $T_{\rm
H^0}^-$ to $T_{\rm eff}$ and neglected the very small ${\rm He^0}$
edge present in these atmospheres.  This simplified representation of
a stellar atmosphere provides enough accuracy to compute the essential
properties of the  emission line spectrum.

We additionally considered an \sed\ derived from the stellar
evolutionary model of Cervi\~no, Mas-Hesse \& Kunth (2004, hereafter
CMK04), which was used by Villar-Mart\'{\i}n et\,al. (2004; VM04) in
their photoionization calculations of the \lan. The selected \sed\
corresponds to a metallicity $\zstar=0.20\,\zsol$ and an age
of 3.4\,Myr.  It is represented by the long-dashed line in
Fig.\,\ref{fig:sed}b.  We included the weak X-ray flux that results
from the conversion of the kinetic energy of the supernova remnants
into X-ray emission (it did not have any effect on the results). The stellar
cluster at that particular age harbours an important population of WR
stars and, as shown by VM04, the resulting ionizing continuum is
sufficiently hard to reproduce the emission line strength of the
\heiiuw\ line observed in the \lan\ spectrum.  
The CMK04 evolutionary models  are characterized by a
powerlaw initial mass function with a Salpeter IMF and stellar
masses comprised in the range 2--120\,\msol.

\section{Model results}\label{sec:resu}

In this section, we present a grid of photoionization calculations for 
comparison with the observed \nchr\ ratios in the two \hzrgs\
and the \lan.  In Sect.\,\ref{sec:aim}, we outline our investigative 
procedure and the format we adopt to display the results. Thereafter, we 
explore the effects of using different \seds\ and varying some of the 
input parameters, as follows: (a) powerlaw photoionization is first studied in
Sect.\,\ref{sec:agn} assuming different metallicities; (b) in
Sect.\,\ref{sec:igm} we study various  \mbr\  energy distributions in
which quasars and stars are contributing in different proportions; (c) 
in Sect.\,\ref{sec:stel} we explore stellar photoionization by metal-free 
atmospheres of varying \teff\ and by a stellar cluster \sed\ containing WR stars.

\subsection{Aims and modelling procedure}\label{sec:aim}

There is a gap of more than four orders of magnitude in the \nchr\
ratio between \onine\ and \otwo. Rather than explain this with a factor 
$\sim 1000$ difference in absorber metallicity between the two \hzrgs\, as in J03, 
we instead explore alternative \seds. As stated in Sec.\,\ref{sec:bck},
our practical goal is to find an \sed\ which reduces the metallicity
gap to $\sim 10$ (that is, obtaining a successful model that use
abundances as low as $\sim 10$\% solar).  We do not aim at obtaining
exact fits of this ratio in each case, but rather at establishing an
order of magnitude agreement between the models and the separate
observations of the thin and thick absorber categories. For this
reason, we only consider the following four widely spaced
metallicities for the haloes: 10, 1, 0.1 and 0.01\,\zsol. The higher
the metallicity, the larger the \nchr\ ratio. The proportionality is
linear except in the high metallicity regime, where the slab
temperature structure is somewhat altered, this effect being more
important in the case of the thick absorbers.  No attempt is made in
this paper to model the background \eelr\ spectrum. The
metallicity of the \eelr\ gas is much higher than (and unrelated to)
the absorber's, as discussed by B00.

The ionization parameter characterizing the models is plotted on the 
abscissa in all figures.  It is a free parameter that cannot be
adequately constrained  with the limited data at hand. 
The target \nchr\ ratio ($y$-axis) for each observational datum is
represented by an horizontal (thick broken) line, since \up\ is not known. Our aim will be to
find models that either cross this observational  line or
come close to it. We consider it unlikely that \up\ is smaller than 0.001,
since C$^{+3}$ would then be reduced to a trace species. 
It is plausible that it takes on much
larger values instead, especially in the case of \otwo\ or the
\lan, since large values of \up\ usually result in larger 
\nchr\ ratios, a characteristic of   these thinner absorbers.  We adopt the 
conservative view that most of the difference between the thin and
thick absorbers may be accounted for by differences in the gas excitation, that is, in \up\ rather
than by metallicity differences only.  Future observations of other
resonance lines might be used to test this (see Sec.\,\ref{sec:cmg}).  

\begin{figure}    
\resizebox{\hsize}{!}{\includegraphics{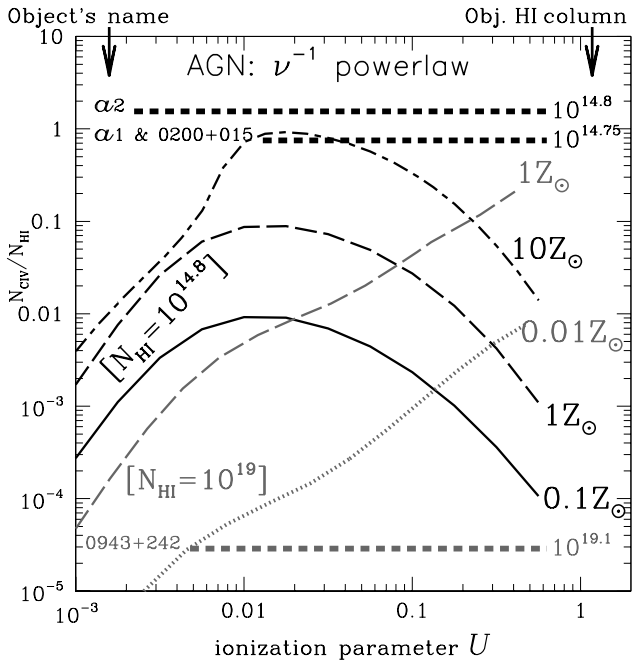}}
\caption[]{The column ratio \nchr\ derived from photoionization by 
a powerlaw of index $-1.0$, as a function of the ionization parameter
\up. Each model sequence (thin lines) along which  \up\ varies
represents a slab of fixed \nhi\ column (shown between brackets). Thin
black lines connect models of constant column $\nhi =
10^{14.8}\,\cms$, while thin gray lines connect models of constant
column $\nhi = 10^{19}\,\cms$. The gas metallicity is shown using
labels, in units of
\zsol. The  {\it thick} horizontal broken lines represent four measurements of
\nchr: the extended absorbers in the high-$z$ radio galaxies \onine\ 
(in gray) and \otwo\ (in black) and the two absorbers
found in the lensed Lynx Arc Nebula [\lan] (in black) and labelled {\it a1} and
{\it a2} by Fosbury et\,al. (2003).  The ratios for \otwo\ and {\it
a1} are very similar and have been combined into a single entry. The
object's name and the
\nhi\ column appear to the left and right, respectively, of the
corresponding horizontal broken line.  }
\label{fig:agn}
\end{figure} 


In what follows, the slab models presented are either characterized by
a small \hi\ column of $10^{14.8}$\,\cms\ as in \otwo\ and the \lan,
or a larger column of $10^{19}$\,\cms\ as in \onine.  
We emphasize that in all the figures (\ref{fig:agn}--\ref{fig:starsb}) of 
Section\,\ref{sec:resu}, the black line models only apply to the thin
absorbers shown at the top while the gray-line models
only apply to \onine\ (shown at the bottom).

\subsection{Powerlaw photoionization}\label{sec:agn}

We present photoionization calculations in Fig.\,\ref{fig:agn} for the
case of an AGN powerlaw of index $-1.0$.  Using a moderately different
index would not significantly alter the conclusions reached below. For
instance, a steeper index $-1.4$ would increase the column ratio only by a
factor of $\la 2$.

In the case of the \onine\ absorber, the models in
Fig.\,\ref{fig:agn} (thin gray lines) favour abundances much
lower than solar, that is of order 1\% solar.  B00 favoured a
metallicity value of $\simeq 0.02\,\zsol$. A much lower (higher) ionization
parameter is a possibility that cannot be ruled out and the
metallicity would then be higher (lower) than the values we
considered.

In the case of the \otwo\ absorber, very high metallicities are
favoured by the powerlaw \sed, as  found by J03. This is
shown by the thin black lines in Fig.\,\ref{fig:agn}, which
suggest a gas metallicity of about ten times solar.  We expressed
concerns about such high values in Sect.\,\ref{sec:bck}. 

We reject the powerlaw \sed\ on account of geometrical considerations
presented in Sect.\,\ref{sec:ion} that led us to rule out direct
ionization by the nuclear radiation from the AGN. The main purpose in
reporting powerlaw calculations is to provide a convenient comparison
with the softer continuum shapes explored below. We note how different the
behaviour of \nchr\ is between the thin and thick absorber case in
Fig.\,\ref{fig:agn} (compare the two long-dashed models of solar
metallicity).

\subsection{Photoionization by the diffuse \mbr}\label{sec:igm}

\subsubsection{\mbr\ flux 
from quasars alone}\label{sec:qso}

\begin{figure}    
\resizebox{\hsize}{!}{\includegraphics{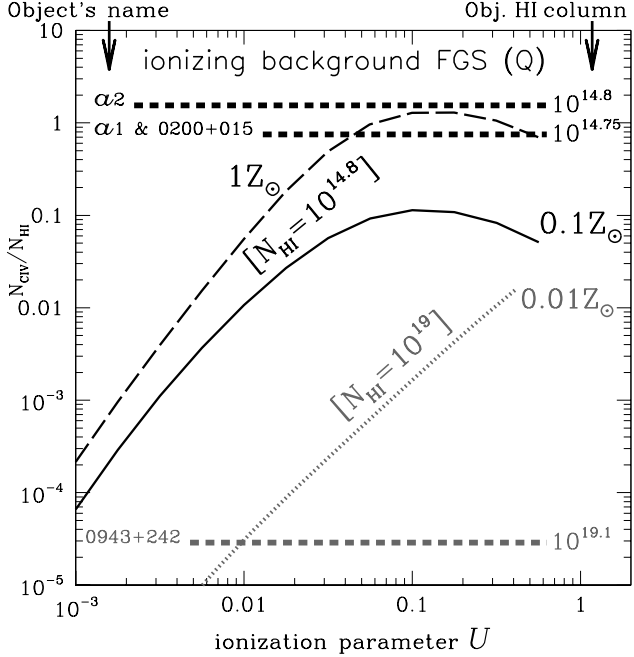}}
\caption[]{The column ratio \nchr\ derived from photoionization  by the  diffuse
\mbr\ due to quasars only, as a function
of \up.  The continuous black line and the long-dashed black line
correspond to the thin absorber case assuming metallicities of 0.1
solar and solar, respectively, while the dotted gray line corresponds
to the thick absorber case assuming metallicities of 0.01 solar.  The
nomenclature and symbols have the same meaning as in
Fig.\,\ref{fig:agn}. In all figures, black line models only apply to
the thin absorber objects shown at the top, while gray-line models
only apply to \onine\ below.}
\label{fig:igmq}
\end{figure} 

The \sed\ of the diffuse \mbr\  resulting from
quasars alone (black long-dashed line in Fig.\,\ref{fig:sed}a) is
significantly softer than a powerlaw. Calculations with such an \sed\ are
shown in Fig.\,\ref{fig:igmq}.  The calculated \nchr\ ratio assuming
10\% solar gas lies below the observed value in \otwo, by a factor
$\sim 10$, as shown by the continuous thin line model in
Fig.\,\ref{fig:igmq}.  Metallicities about solar would be required so  that the model overlaps 
the observed column ratio\footnote{Hence
in the thin slab case, the increase  in \nchr\ provided by a quasar-based
\mbr\ \sed\  is a factor ten with respect
to a straight AGN powerlaw. }. As for \onine, in absence of definite
information about \up, the absorber's metallicity cannot be
constrained any further than in the previous powerlaw case covered in
Sect.\,\ref{sec:agn}.

\subsubsection{\mbr\ flux from stars and quasars}\label{sec:fgs}

\begin{figure}    
\resizebox{\hsize}{!}{\includegraphics{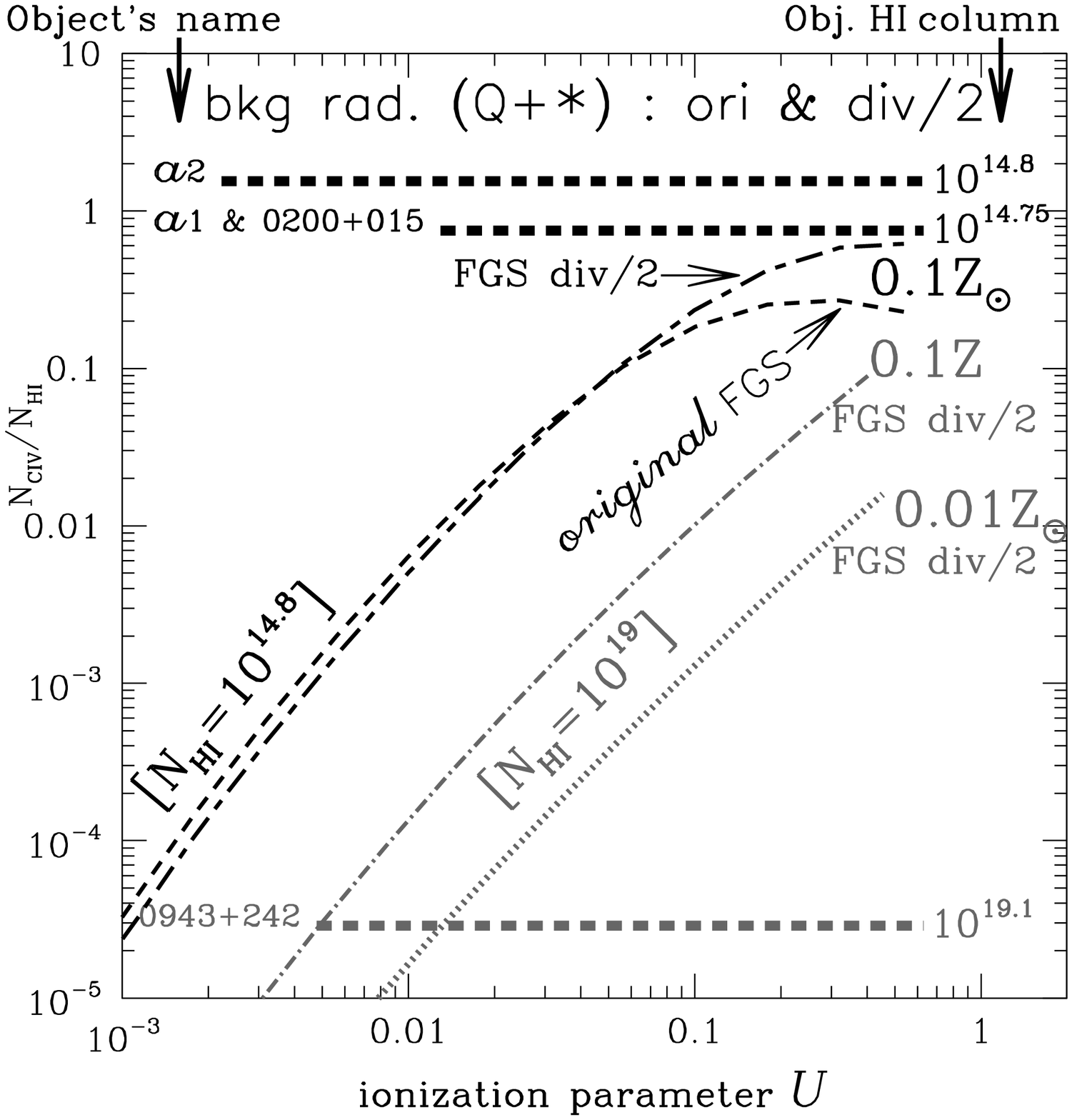}}
\caption[]{The column ratio \nchr\ derived from photoionization  
by the  \mbr\ due to stars {\it and} quasars, as a function of
\up.  The black short-dashed line corresponds to photoionization by the 
{\it original} FGS \sed\ in which the \mbr\ flux is
generated by quasars as well as stars, assuming a gas metallicity of
0.10\,\zsol. The black short-long dashed line has an \sed\ similar to
the previous, but its flux beyond 54\,eV has been halved (the
dotted-line \sed\ in Fig.\,\ref{fig:sed}a), owing to a larger
contribution by stars.  The gray dot-dashed line and gray dotted line
assume this latter \sed\ in the thick  slab case, but with
metallicities of 0.1 and 0.01\,\zsol, respectively. The nomenclature
and symbols have the same meaning as in Fig.\,\ref{fig:agn}. }
\label{fig:igmqst}
\end{figure} 

In the case in which stars and not just quasars contribute to the
\mbr, the ionizing \sed\ becomes softer (continuous line in
Fig.\,\ref{fig:sed}a) and the \nchr\ ratio observed in \otwo\ can now
be reproduced using metallicities not much above 10\% solar, as
illustrated by the thin dashed line in Fig.\,\ref{fig:igmqst}. The
short-long dashed thin line in Fig.\,\ref{fig:igmqst} represents the
case of an even softer
\sed\ in which the flux beyond 54\,eV has been halved (see dotted line
\sed\ in Fig.\,\ref{fig:sed}a). This latter \sed\  hence satisfies our initial goal 
defined in \ Sect.\,\ref{sec:aim} with respect to metallicity. But
adopting an \sed\ in which star-forming galaxies are contributing more
than quasars does not imply that such a distribution is typical of the 
{\it average} \mbr. It only suggests that 
this \sed\ is valid  in the neighborhood of
\otwo. Smette et al. (2002) found for instance that the softness of
the \mbr\ energy distribution (parameter \etap, see
Sect.\,\ref{sec:soft}) presents important local variations, with some
locations where only quasars are apparently contributing while in
others there appears to be a significant contribution from starbursting
galaxies.

In the case of the thicker absorber in \onine, there is little
difference in \nchr\ between the Q+{\Large\mbox{$\star$}} \sed\ in which the
flux beyond 54\,eV has been halved and the \sed\ produced
by quasars only. Compare, for instance, the gray dotted lines
(0.01\,\zsol) in Figs\,\ref{fig:igmq} and \ref{fig:igmqst}.

In summary, a diffuse \mbr\ sustained by quasars
and star-forming galaxies is quite successful in reproducing the
observed column ratio in \otwo\ without need for a metallicity any
higher than $\sim 0.10$\,\zsol.

\subsection{Photoionization by local stellar UV}\label{sec:stel}

\subsubsection{Photoionization by hot stars with \teff=80\,000\,K}\label{sec:hot}

Using a metal-free stellar atmosphere of 80\,000\,K and a gas
metallicity of 4\% solar, BG03 obtained a reasonable first order fit to
the strong lines observed in the unusual spectrum of the high redshift \lan.  
In Fig.\,\ref{fig:starsa}, 
we show that the column density ratios of the \lan\ absorbers 
can be reproduced using a high value of \up\ and an absorption gas metallicity of 0.1--0.2\,\zsol. 
This range
is consistent with the comprehensive metallicity determination of the
nebular emission gas by VM04, that is $\simeq 10$\% solar. Given the similarity
of the \lan\ column ratio with that of \otwo, we infer that a
\nchr\ ratio of order unity in an \hzrg\ is compatible with  a
stellar \sed\ photoionizing a subsolar metallicity absorber.  Hence, the
possibility that the \otwo\ absorber might be photoionized by hot
stars warrants consideration, since metallicities of only 10\% solar
would be needed rather than a value 100 times larger favoured by the
powerlaw \sed\ (Fig.\,\ref{fig:agn} or J03). The problem of 
stellar continuum detection is discussed in Sect.\,\ref{sec:weak}.

\begin{figure}    
\resizebox{\hsize}{!}{\includegraphics{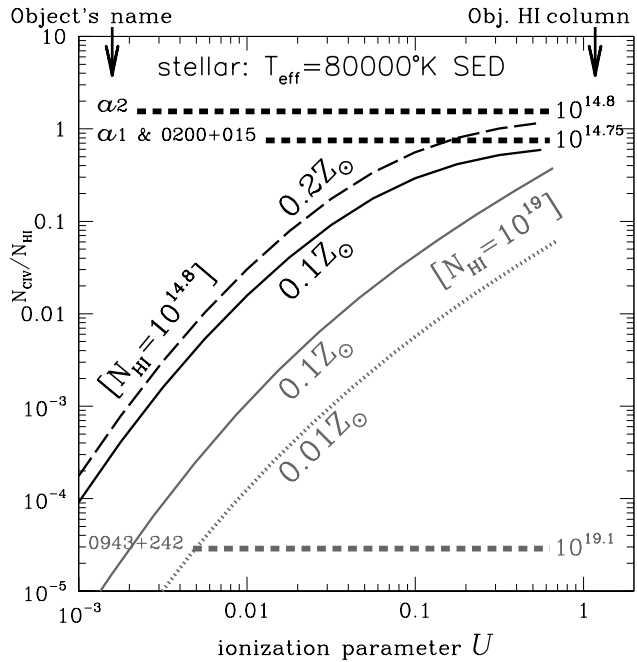}}
\caption[]{The column ratio \nchr\ derived from 
photoionization  by  a metal-free stellar  atmosphere  
 with \teff=80\,000\,K as described in Sect.\,\ref{sec:str}, as a
 function of \up. The nomenclature and symbols have the same meaning
 as in Fig.\,\ref{fig:agn}.}
\label{fig:starsa}
\end{figure} 

In the case of thicker absorbers, as represented by \onine, the
gray line models point to metallicities in the range 0.01--0.1\,\zsol,
assuming $\up < 0.05$. Higher values of \up\ would imply lower absorber
metallicities.  Interestingly, the stellar (Fig.\,\ref{fig:starsa})
and the powerlaw (Fig.\,\ref{fig:agn}) models with 0.01\,\zsol\ cross
the \nchr\ ratio of \onine\ at a very similar \up\ value. This
indicates that thicker slabs are much less sensitive to the \sed's
shape.


\subsubsection{Varying the stellar atmosphere temperature}\label{sec:var}

\begin{figure}    
\resizebox{\hsize}{!}{\includegraphics{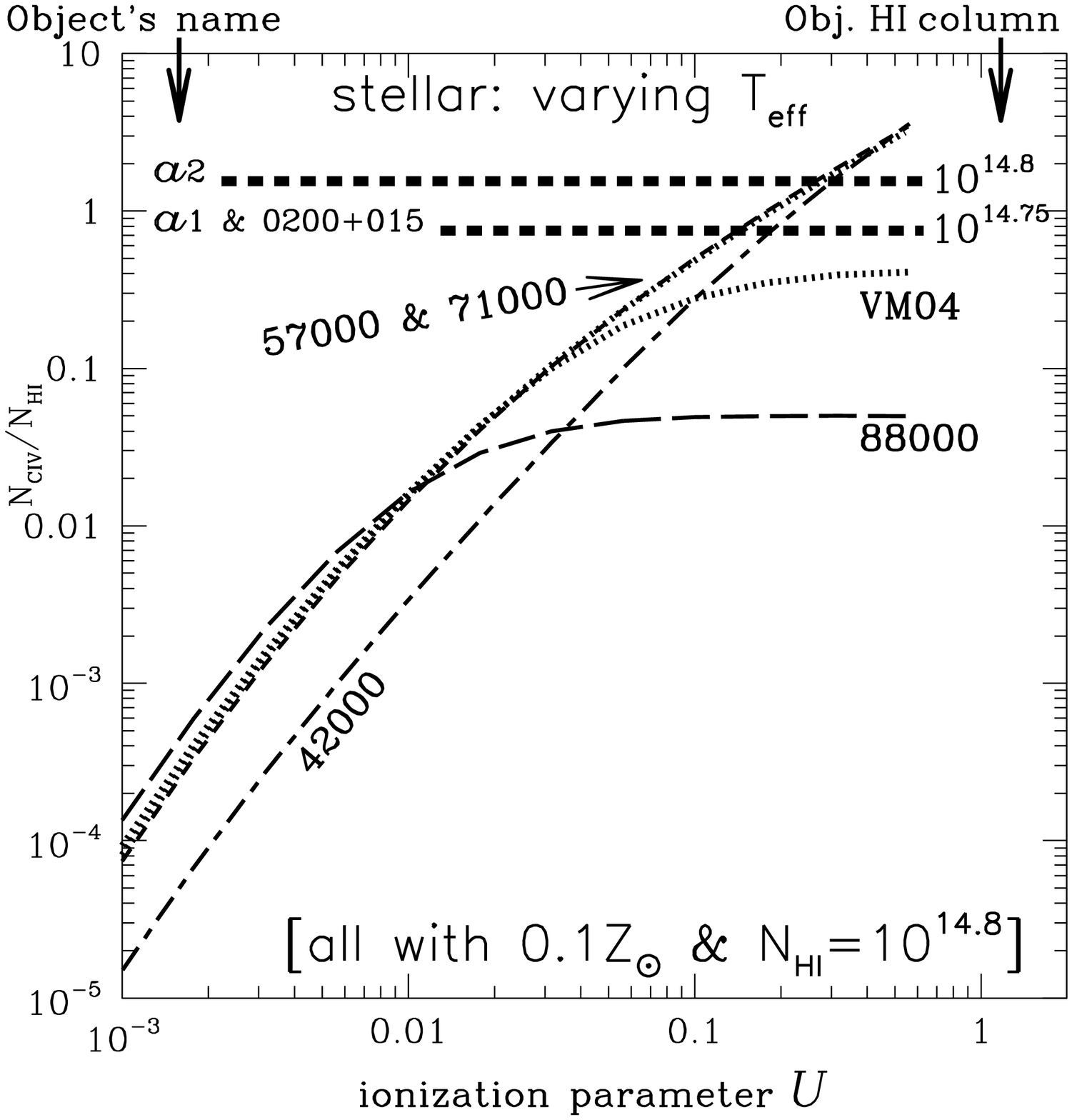}}
\caption[]{The column ratio \nchr\ derived from 
photoionization by stellar energy distributions of varying \teff, as a
function of \up. In all models, the metallicity is 0.1\,\zsol\ and the
slab opacity  $10^{14.8}$\,\cms. A label shows the \teff\ being
considered.  The dotted line labelled VM04 corresponds to
photoionization by a stellar cluster of metallicity 20\%  solar and
3.4\,Myr of age, as described in Sect.\,\ref{sec:str}. The
nomenclature and symbols have the same meaning as in
Fig.\,\ref{fig:agn}. }
\label{fig:starsb}
\end{figure} 
  
In Fig.\,\ref{fig:starsb} we explore the effect of varying stellar
effective temperature. The zero-age metal-free atmospheres used
correspond to \teff\ of 42\,000, 57\,000, 71\,000, 80\,000 and
88\,000\,K (from Schaerer 2002). All models are characterized by a
column $\nhi = 10^{14.8}\,\cms$ and a gas metallicity of 10\% solar,
appropriate to the \lan.  Although temperatures lower than 80\,000\,K
can easily fit the \lan\ column ratio, this would imply too weak
\heiiw\ emission for the nebula.  At the other temperature end, a
\teff\ as high as 88\,000\,K would require a ten times higher gas
metallicity in order to reproduce the observed column ratio. The
reason is that, as \teff\ is increased much beyond 70\,000\,K, the
increase in the continuum's hardness causes the slab to harbour many
ionization stages of carbon (e.g. C$^{+4}$ and C$^{+5}$), thereby
causing a relative reduction of the \civ\ fraction. Increasing the
temperature much beyond $10^5\,$K would result in column ratios
approaching those of a powerlaw.

VM04 have modelled the \lan\ emission line spectrum using the ionizing
spectrum of an evolved stellar cluster in which transient Wolf-Rayet
stars can account for the nebular \heiiw\ line observed in emission. A
photoionization model of the absorber using such an \sed\ is
represented by the dotted line labelled VM04 in
Fig.\,\ref{fig:starsb}. The  behaviour of the column ratio is
similar to that of a 80\,000\,K metal-free star (Fig.\,\ref{fig:starsa}).

In the case of the \hzrg\ shells, stellar \seds\ are possible
candidates for the ionization of the absorbers (but not of their
\eelr), since they can reproduce the observed column ratio using subsolar metallicities. 

\section{Estimates of the radii and masses of the shells and compatibility with other observables}\label{sec:esti}

Armed with the results of the photoionization calculations, we now investigate 
two of the scenarios in more detail, namely the cases of ionization by the 
\mbr\ and by hot stars. We focus on their implications for other properties of 
the absorbing shells (e.g. mass, radius and thickness) and their compatibility 
with other observables (e.g. the underlying stellar continuum in the case of ionization by hot stars, and the strength of the \lya\ emission). On the basis of such considerations we demonstrate that ionization by hot stars is favoured over \mbr\ ionization. Readers who do not wish to follow the argument in full may skip over sections \ref{sec:sizm} and \ref{sec:sizs} and proceed directly to the summary in section \ref{sec:summ}.

\subsection{The case of \mbr\ ionization} \label{sec:sizm}

We first analyse the possibility of having the \mbr\ ionize the
haloes.  The diffuse \mbr\ is ubiquitous and its intensity independent of
distance to the \hzrg. Therefore, changes in
excitation (i.e. \up\,) are obtained by varying the gas
density. Higher shell densities, hence lower \up, might explain the 
absence of \civ\ absorption in many \hzrg\ shells.
We must thus investigate whether the \mbr\ is strong enough to result in an acceptable
halo density, because the geometrical thickness of the shell increases as the
density is reduced. Below we use such constraints to infer the shell's minimum distance 
from the \hzrg\ and its total mass.

\subsubsection{\mbr\ intensity and shell thickness}\label{sec:jnu}

An estimate of the \mbr\ mean intensity is provided by the proximity effect, 
whereby absorbers becoming more ionized in the vicinity of quasars.  We will adopt the
value $\jnu \approx 10^{-21}$\,\inu\ inferred by Cooke, Espey \&
Carswell (1997) and assume the \mbr\ flux \sed\
Q+{\Large\mbox{$\star$}} of Fardal et\,al. (1998), albeit with the flux
above 4\,Ry divided by two, as studied in Sect.\,\ref{sec:fgs}. We
will consider two cases: the weak and the strong absorber cases, using
\otwo\ and \onine\ as examples, respectively. Using the
definition\footnote{If the \sed\ was a powerlaw of index $\alpha$
($<0$), we would have $\up \approx -4 \pi \jnu /(\nnh h c \alpha)=
6.3\,10^{-5} \jto /n_{\rm H}$, where $h$ is the Planck constant and
$c$ the speed of light.} of \up\ and \map\ to integrate the
Q+{\Large\mbox{$\star$}} \sed, we obtain in the optically thin case
that the {\it total} hydrogen density is given by  $\nn = 2.8 \times
10^{-4} \jto \{\upn\}^{-1} \cmc$, where $\upn = U^{\rm thin}/0.1$ and
$\jto= \jnu/(10^{-21} \inu)$. In the case of thicker absorbers with
$\nh \ga 10^{18}\,$\cms, because of self-shielding, illumination of a
spherical shell can only occur from the outside, and the mean
intensity is approximately half of the previous thin case, such that
$\nk = 2.8 \times 10^{-3} \jto \{\upk\}^{-1} \cmc$, where (for
convenience) $\upk = U^{\rm thick}/0.005$.  The photoionization
calculations at constant \nhi\ (either $10^{14.8}$ or $10^{19}$\,\cms)
indicate that the {\it total} hydrogen column of the slab can be
approximated as $\nhn \simeq 2.1 \times 10^{19} \, \{\upn\}^{1.1}$\,\cms\
and $\nhk \simeq 6.3 \times 10^{20} \, \{\upk\}^{1.1}\,$\,\cms,
respectively. As argued in Sect.\,\ref{sec:geo}, a shell geometry is
more appropriate than that of a filled sphere. We therefore introduce
an aspect ratio $A = \Delta r/r$ for the shell, where $\Delta r$ is
the shell thickness and $r$ its outer radius, taking the \hzrg\
nucleus as the center. Since this ratio is not known, we will define
an upper limit of $A \la 0.2$.  Since the {\it total} column density
is given by $\nh = r n_{\rm H} A$, this limit on $A$ translates into a
lower limit for the shell radius (i.e.  a minimum radius) of $\rn
\ge 122 \, \{\upn\}^{2.1} \{\at \jto\}^{-1}$\,kpc and $\rk \ge 365 \, \{\upk\}^{2.1} \{\at
\jto\}^{-1}$\,kpc, respectively, with $\at = A/0.2$.  To be definite, we will assume that both
absorbers have the same gas metallicity of 0.10\,\zsol. From
Fig.\,\ref{fig:igmqst}, we read off values of $\upn \simeq 2$ and
$\upk \simeq 1$ for \otwo\ and \onine, respectively. This translates
into minimum radii of 523 and 365\,kpc, respectively.  Hence \mbr\
ionization implies that the shells are extremely distant from the
background \eelr,  and this appears to be difficult to reconcile with
the observations which show a distinct transition from sources with
radio extents  $< 25$~kpc to those with large radio sizes (see 
section~\ref{sec:test}).

Since $r_{\rm kpc} \propto U^{2.1}$, smaller radii follow from assuming 
smaller values of \up, which would require
that we adopt metallicities somewhat larger than 0.1\,\zsol\ (larger metallicities  shift  
models to  the left in Fig.\,\ref{fig:igmqst}).
Uncertainties in \up\ (or equivalently in $Z$)  therefore affect
 our estimates of the shell's geometrical thickness significantly. In
Sect.\,\ref{sec:cmg} and \ref{sec:on}, we indicate how detection of
the shells in \mgii\ or \ovi\ would help to constrain both $Z$ and \up.

\subsubsection{Problem of the large shell masses}\label{sec:mass}

Since the shell masses are given by $\mh = 4 \pi r^2 m_{\rm H} N_{\rm
H} = 10^{-13} \, r^2_{\rm kpc} N_{\rm H}$\,\msol, assuming they are
spherical, we can use the previous expressions for the minimum radii
to derive the following minimum masses $\mhn \ge 3.1 \times 10^{10} \,
\{\upn\}^{5.3} \{\at \jto\}^{-2}\,$\msol\ and $\mhk \ge 8.3 \times
10^{12} \, \{\upk\}^{5.3} \{\at \jto\}^{-2}\,$\msol, respectively.
Adopting the same estimates of \up\ as above, we derive masses of
$\mhn \ge 1.2 \times 10^{12}\,$\msol\ and $\mhk \ge 8.3 \times 10^{12}
\,$\msol\ for \otwo\ and \onine, respectively.  At face values, these values 
are excessive and would suggest that \mbr\ ionization is unworkable. On the other hand,
given the strong dependence of \mh\ above on poorly determined
quantities, the upper limits mentioned above are order of magnitude estimates 
and as such do not allow us to completely rule out \mbr\ ionization.  For instance, 
reducing both ionization parameters by two reduces the
\otwo\ and \onine\ shell mass estimates to $3.1 \times 10^{10}$ and
$2.1 \times 10^{11}\,$\msol, respectively.
Proportionally smaller masses would be implied if the
shells covered only a fraction of $4\pi$\,sr.  On
the other hand, if the shells were to be geometrically very thin
(i.e. $\at \ll 0.2$), the mass of the \onine\ shell would become
unreasonably large ($> 10^{13}$\,\msol).

In conclusion, the ionization of the shells by the diffuse \mbr\ would
imply that the shells have expanded to large distances from the
parent \hzrg.  This would favor the `aging shell' scenario B. A
significant problem is that the shell mass estimates  turn out too large.  
An alternative is that the ionizing radiation is stronger as a
result of {\it local} stellar sources, as discussed below.

\subsection{The case of ionization by local stellar sources} \label{sec:sizs}

The similarity of the \nchr\ ratio between the stellar-excited \lan\ nebula and the
\otwo\  absorber suggests that hot stars could be the ionization source of the 
\hzrgs\ haloes.  Even though the emission-line spectra of the
background \eelr\ is clearly AGN-like and presumably ionized by the
hidden quasar, an interesting result of the calculations in
Sect.\,\ref{sec:stel} is that the column ratios in the two \hzrg\ {\it absorbers} can
be reproduced using a stellar \sed\ and subsolar metallicites, as for
the \lan.  We now analyse some of the implications of this hypothesis.

\subsubsection{Test case: hot stars contributing little  to the \eelr}\label{sec:test}

The geometry that we envisage is that of a large population of hot
stars, possibly distributed uniformly or in large aggregates as a
result of merging (e.g. the \hzrg\ 4C\,41.17; van\,Breugel et\,al. 1997). 
To simplify the treatment of the geometrical dilution of the ionizing radiation, 
we will assume
that the propagation of the photons is approximately radial by the
time they reach the intervening shells. To facilitate the comparison
with the previous \mbr\ ionization case, we define a reference {\it
test case} with a much higher shell density of
0.01\,\cmc.  The photon density is set by the relation $\nnh = 10^{-2}
\, \{\upvv\}^{-1} \cmc$, which is equivalent to having 
an ionizing flux  (reaching the shell) 36 times higher than that provided by
the \mbr\ intensity with $\jto = 1$, as assumed in
Sect.\,\ref{sec:jnu}. Under the conditions of this test
case, our calculations indicate that the ionizing photon flux
impinging upon the inner boundary of the shells is $\phih = 3.0 \times 10^{7}
\, \nvv \upvv$\,\qtscm, where $\nvv =\nnh/0.01$ represents the shell
density.  Local stellar sources (in contrast to the \mbr\ 
case) are in better accord with the `inner shell' scenario A, in which the shells
do not extend further out than about 25\,kpc in radius, i.e. the
apparent crossover point between sources with absorbers and those without [see
J03; W04;  Sect.\,\ref{sec:shel} and the superwind-bowshock model of Krause
(2005)].  The photon luminosity is $4 \pi r^2 \phih$, which can be
written as $\qh= 0.224 \, 10^{55}\, r^2_{\rm 25} \,
\nvv \, \upvv $\,\qts, where $\rt = r_{\rm kpc}/25$. The \lya\
luminosity from recombination alone is given by the expression $\Lya =
1.06\times 10^{-11} \epsn \, \qh$\,\ergs, where the conversion
factor\footnote{Under the quoted physical conditions, 65\% of the
recombinations lead to \lya\ photon emission (e.g. Binette
et\,al. 1993).} assumes case\,B and a temperature of 20\,000\,K.
\epsn\ is the fraction of photons absorbed and reprocessed by the
emission gas.  The leaking fraction $1-\epsn$ for very luminous \hii\
regions lies in the range 0.3--0.5 (Beckman
et\,al. 2000 and references therein; Zurita et.\,al. 2002; Rela\~no
et\,al. 2002; Giamanco et\,al. 2005).  
To be definite, we adopt 0.5 and define $\epsnt=
\epsn/0.5$ to obtain that 
\begin{eqnarray} 
\Lya = 0.12 \times 10^{44}\,\epsnt \, r^2_{\rm 25} \, \nvv \, \upvv \,\ergs\ \label{eqn:str}
\end{eqnarray}
for our test case.

The \Lya\ luminosity in the test case should be compared with the
significantly larger \eelr\ \Lya\ luminosities of $1.2 \times 10^{44}$ and
$1.9 \times 10^{44}$\,\ergs, observed in \otwo\ and \onine,
respectively. In the case of the \lan, with  $\Lya =
0.40 \times 10^{44}$\,\ergs, its luminosity\footnote{The  observed
\Lya\  values quoted above were derived using the \lya\ fluxes
reported by VO97 and that of Fosbury et\, al. (2003) for the \lan. We
assumed the objects to be isotropic emitters and corrected the fluxes
for \lya\ absorption due to the absorbing shells. For the \lan,  \Lya\ 
was divided by 10 to compensate for the amplification by the
gravitational lens.  We adopted the concordance $\Lambda$CDM cosmology
with parameters with $\Omega_{\Lambda} = 0.7$, $\Omega_{M}=0.3$, $h =
0.70$ with $h = H_0/100$. The isotropic photon luminosities that we infer for
\otwo, \onine\ and the \lan\ are $\qh = 1.1 \times 10^{55}$, $1.8 \times 10^{55}$ and
$0.37 \times 10^{55}\,$\qts, respectively. These values are lower limits,
since they only represent the fraction absorbed by the gas and
reprocessed into line emission.  To recover the intrinsic
\qh, they would have to be increased by $\epsn^{-1}$,
a poorly determined quantity in AGN ($\epsnagn \la 0.1$ : Oke \&
Korycansky 1982; Antonucci et\,al. 1989).}  is three times higher than
our test case.  As for our two \eelrs, they are brighter in \lya\ by a
factor 5 (\otwo) and 300 (\onine), assuming in Eqn.\,1
that $\upn \simeq 2$ and $\upk \simeq 1$, respectively. The assumed
stellar ionizing luminosity is therefore not expected to alter the AGN
character of the \hzrg\ emission spectrum, even though specific
emission lines would be subject to a contribution from the proposed
stellar sources.

Interestingly, the  \lya\ luminosities of the absorption shells
themselves are expected to be relatively small.  We derive \lya\
luminosities of $\Lya = 5.5 \times 10^{43}\, \epsh \, \nvv
\, \upvv$\ergs, where  \epsh\  is the fraction of ionizing photons 
absorbed by the shell, a quantity that is set by the shell
opacity.  Our calculations with \nhi\ of $10^{14.8}$ and
$10^{19}$\,\cms\ indicate that $\epsh = 5 \times 10^{-4}$ and 0.97 for the
thin and thick absorbers, respectively. Assuming as in
Sect.\,\ref{sec:jnu} that $\upn \simeq 2$ and $\upk \simeq 1$ for
\otwo\ and \onine, respectively, this translates into luminosities of
$5.5 \times 10^{40}$ and $2.7 \times 10^{41}$\,\ergs, respectively. These values are
negligible with respect to the observed \hzrg\ and \lan\ \lya\ luminosities. 

As for the masses of the shells, the total column densities as a
function of \up\ in the case of stellar \sed\ are as follows: $\nhn
\simeq 1.1 \times 10^{19} \, \{\upn\}^{1.1}$\,\cms\ and $\nhk \simeq 7.2 \times 10^{20} \, \{\upk\}^{1.1}\,$\,\cms, assuming the 80\,000\,K
\sed. We used the expression $\mh = 6.25 \times 10^{-11} \, r^2_{\rm 25}
N_{\rm H}$\,\msol, assuming again that the shells are spherical, to
derive mass estimates of $\mhn = 6.7\times 10^{8} \, r^2_{\rm 25} \,
\{\upn\}^{1.1}$\,\msol\ and $\mhk = 4.5 \times 10^{10} \, r^2_{\rm 25}
\, \{\upk\}^{1.1}$\,\msol\ for \otwo\ and \onine, respectively. 
Hence, the shell masses for the test case are quite small in
comparison with the \mbr\ case because of the smaller radii
implied by the stronger ionizing flux.  By the same token,
larger densities are implied, which result in shells that are also
geometrically very thin ($\at \ll 1$).  The absorber's density can be
quite different than the assumed test case with $\nvv=1$. The required
stellar luminosity, however, must then scale in the same
proportion. For instance, an absorber with density 0.1\,\cmc\ would
require a 10 times higher stellar luminosity. This would cause the
nebular lines to be comparable in luminosity to the observed \eelr\, which would
clearly not be desirable.

\subsubsection{On the detection of stellar continuum}\label{sec:weak}

For the case where hot stars alone ionize the foreground absorbers, we now
estimate the implied stellar flux (or, equivalently, the \lya\ equivalent-width)
and compare it with the observations, beginning with the \lan. 

Assuming a Salpeter IMF and the 
\sed\ for an instantaneous burst of age 3.4\,Myr (VM04), 
we find using \map\ that the rest-frame \lya\ equivalent-width is
$\ewr = 190 \,\epsn$\,\AA. Defining the fraction of ionising photons 
reprocessed by the emission nebula as $\epsnt=
\epsn/0.5$, we obtain that the (observer-frame)  continuum flux   is 
$\fco= 0.0105  \, \{\epsnt \, (1+\ze)\}^{-1} $\,\flyo\AA$^{-1}$, 
where \flyo\ is the observed line flux,  corrected
for absorption. 

For the \lan, this implies a 5300\,\AA\ continuum of $\fco = 9.4 \times 10^{-18}\,$\fla, 
or equivalently, 8.8\,\mjy. (The lens amplification was assumed to be
the same for both the continuum and the lines). This flux is about 30
times higher than the upper limit set by Fig.\,5 of Fosbury
et\,al. (2003), of $\approx 0.3$\mjy.  A continuum was detected at
longer wavelengths, but these authors report that it is consistent
with being nebular in nature.  As a solution, Fosbury et\,al. (2003)
proposed a top-heavy IMF.  Assuming a single \teff\ \sed\ of
80\,000\,K (Fig.\,\ref{fig:sed}), we derive $\ewr = 1335\,\epsn$\,\AA,
or a continuum of 1.3\,\mjy ($\epsnt=1$).  Even for a higher \teff\ of
88\,000\,K, we obtain a value of 1.1\,\mjy, similar to before. The
increase of a factor 7--8 of the
\lya\ equivalent-widths provided by these two \seds\ is therefore
insufficient.
A significantly hotter stellar \sed\ is therefore required (Fosbury
et\,al. proposed $\simeq 10^5\,$K).  Alternative explanations might
consist of a peculiar dust distribution that preferentially absorbs
the continuum and thereby increases the observed equivalent-width or,
as suggested by MV04, differential amplification of the lines and the
continuum, by the gravitational lens. It is interesting to note that
the \lya-emitting `blobs' associated with Lyman Break Galaxies
likewise do not show the expected level for the stellar continuum
(Steidel et\,al. 2000).  For a possible explanation involving significant
populations of metal-free stars, see Jimenez \& Haiman (2006). 

If we now turn to the two \hzrgs, we place upper
limits on the underlying stellar continua by defining limits
on the contribution of hot stars to the \eelr\ \lya\ (which must be much smaller
than the AGN contribution). Assuming the VM04 \sed\ and the stellar
contribution to be no more than 10\% of the \eelr, we derive
continuum fluxes (at the observed \lya\ wavelength) of $7.0 \times
10^{-19}$ and $1.0 \times 10^{-18} $\,\fla, for \onine\ and
\otwo, respectively\footnote{It should be emphasized that these are  
maximum estimates of the stellar continuum.  We recall that we can
let the stellar continuum be much {\it weaker} than the test case
explored in Sect.\,\ref{sec:test} and still have it ionize the
shells. Furthermore, if $\up \ll 0.1$, as considered for \otwo\ in
Sect.\,\ref{sec:hot}, an even weaker continuum is needed, as shown by
the estimates of \lya\ luminosity reported in Sect.\,\ref{sec:weak}.}.
These lie below the upper continuum limits of $3.1\times 10^{-18}$ and
$2.6 \times 10^{-18}$\,\fla, respectively, as measured by van\,Ojik (1995).
However, using recent VLT data from VIMOS-IFU (van Breukelen, Jarvis
\& Venemans 2005), one of us (MJ) reports detection of the underlying continuum in
\onine\ at the level of $1.7\pm{0.9} \times 10^{-18}$\,\fla, that is, less
than a factor two above our limit.  Vernet et\,al. (2001) reports on
the measurement of a far-UV continuum in the form of ``single peaked
sources''. This continuum, however, is 6.6\% polarized near
1350\,\AA.  Vernet et\,al. (2001) estimates that the AGN contributes
between 27\% and 66\% of the continuum at 1500\,\AA.  After allowing
for a 20\% contribution from the nebular continuum, these authors
conclude that between 14 and 55\% of the unpolarized continuum might
be due to young stars.  The continuum measured by MJ is then
fortuitously consistent with our upper limit, since half of it or less
is stellar in origin.  We conclude that an instantaneous burst with a
Salpeter IMF is thus a feasible source of ionization for \hzrg\
absorbers. It would in any case be difficult to rule it out since continua much 
weaker than assumed in our test case (by a factor $\sim 20$) would still suffice 
to ionize the absorbers.



\subsection{Summary of current constraints on \hzrg\ halo ionization} \label{sec:summ}

To summarise the results of the previous two sections, we conclude
that ionization by the \mbr\ or by hot stars can satisfactorily reproduce 
the observed \nchr\ ratios without recourse to excessive galaxy-to-galaxy
metallicity variations. That was the aim of the photoionization modelling 
as defined in section \ref{sec:aim}. 

On closer inspection, however, ionization by the \mbr\ leads to excessively 
large radii for the absorbing shells and, by implication, to very large gas masses. 
This follows because the intensity of the \mbr, \jnu, is not a free parameter so 
constraints on \up\ translate directly into constraints on halo
gas density. Given the latter, the observed requirement for a shell-like geometry 
translates directly into a minimum shell radius from the parent \hzrg. For 
both thick and thin absorbers, the minimum radii are of order several hundred kpc. 
This is hard to 
reconcile with the observed transition in radio source size between \hzrgs\ with 
and without strong absorption. Scaling as the square of the radius, the 
implied shell masses are also uncomfortably large.

The case of ionizing the absorber, but not the \eelr, by hot stars circumvents 
the above problems, but at first sight raises separate issues of its own. The 
first is to ensure that these hot stars do not overproduce the \lya\ 
emission, because in \hzrgs\ the \eelr\ is powered by AGN photoionization or 
jet interactions. In both \otwo\ and \onine\, it was shown that \lya\ emission from
hot stars does not significantly contaminate the \eelr\ emission. Potentially 
more serious is the apparent faintness of the stellar continuum, which
is hard to explain away with peculiar dust geometries if the stars ionize gas in the direction of the observer. For the \lan\, Fosbury et al. (2003) appealed to hot stars and a top-heavy IMF; for the two \hzrgs, constraints on the continuum 
level below \lya\ appear to be consistent with the levels expected from hot stars. For all these reasons, we thus favour hot stars as the more likely source of ionization for the \hzrg\ haloes and in the next section outline some new diagnostics to test this further.

\begin{figure}    
\resizebox{\hsize}{!}{\includegraphics{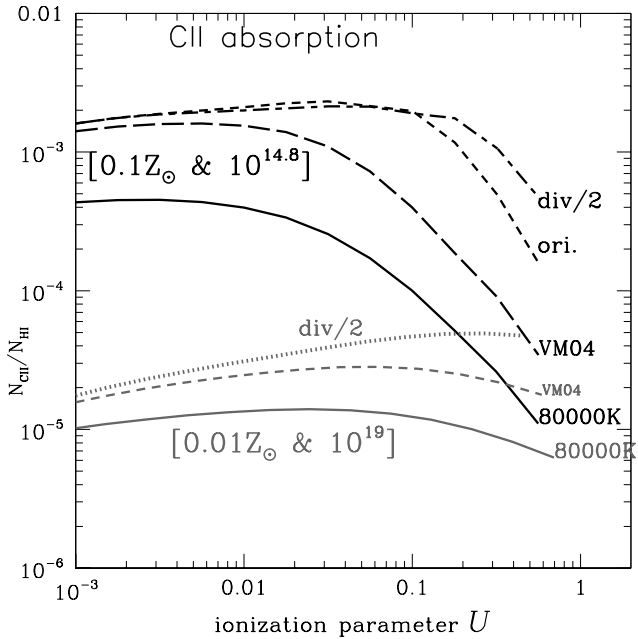}}
\caption[]{The column ratio \nchrl\ derived from photoionization  by  four different  
\seds\ discussed  in Sect.\,\ref{sec:resu}, as a function of \up. The nomenclature and
symbols have the same meaning as in Fig.\,\ref{fig:agn}.
}
\label{fig:c2}
\end{figure} 

\begin{figure}    
\resizebox{\hsize}{!}{\includegraphics{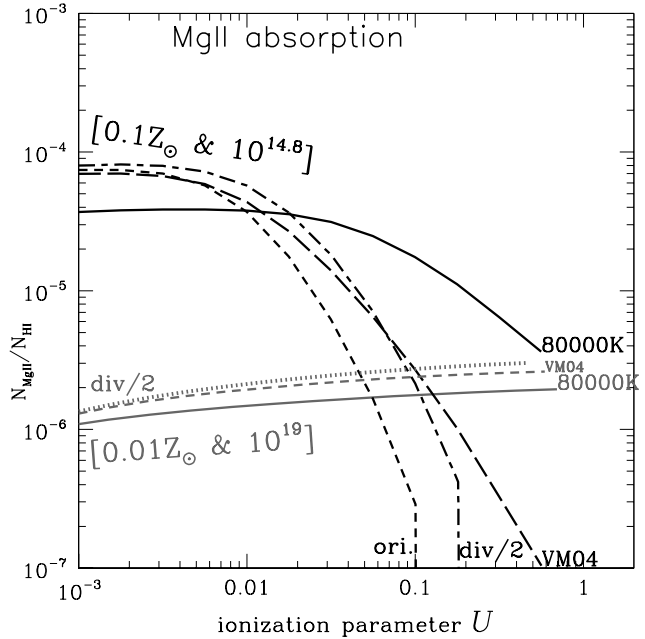}}
\caption[]{The column ratio \nmghr\ derived from photoionization  by  four  different 
\seds\ discussed  in Sect.\,\ref{sec:resu}, as a function of \up. }
\label{fig:mg2}
\end{figure}


\section{New diagnostics for future observations}\label{sec:futu}

In order to resolve pending issues such as the size, masses and nature
of the \hzrg\ haloes, more extensive observations are needed and
measurements of \civ\ in absorption in other \hzrgs\ should be
attempted. The detection of other absorption species would also  help to
break the $Z$--\up\ degeneracy, as outlined below. 

\subsection{Absorption by lower ionization 
species: \ciiw\ and \mgiiw}\label{sec:cmg}

It would be helpful to detect the absorption of  other
resonance lines in the spectra of \hzrgs, particularly in the case of species of lower
ionization than \civ. This could be used to confirm  whether those
\hi\ absorbers without \civ\ absorption  might simply correspond to shells of
lower ionization (smaller \up\,).  Two candidate species are \ciiw\
and \mgiiw. We report calculations for these two resonance lines in
Figs\, \ref{fig:c2} and \ref{fig:mg2}, assuming those  \seds\
that were most successful in reproducing the observed
\nchr\ ratios.  Because both \cii\ and \mgii\  are much weaker {\it emission} lines  than
\lya,   we consider it feasible to detect
the corresponding absorption doublets  only  in the case of the thicker \hi\ absorbers. 
For instance, for an absorber with $\nhi \simeq 10^{19}\,$\cms, we expect the
\cii\ and \mgii\ columns to be of order $10^{14}$ and
$10^{13}\,$\cms, respectively, assuming a  metallicity of
0.01\,\zsol.  Interestingly, the behaviour of the \nchrl\ and
\nmghr\ ratios is relatively flat in the strong absorber case, with a  dependence on
\up\ that is much weaker than was the case for \civ.  
This property would facilitate the determination of the gas
metallicity.  A possible strategy would be to use \mgii\ to ascertain
the metallicity, and then use the appropriate \civ\ curve to constrain  \up.

\subsection{Absorption by higher ionization species: \oviw\ and \nvw}\label{sec:on}

McCarthy (1993) produced a composite
optical-UV spectrum of 3CR and 1\,Jy sources (redshifts up to 3) that
is useful for estimating  typical strengths of various emission
lines. Their composite shows that the strongest resonance emission
lines in radio-galaxies after \lya\ and \civw\ are (in order of
decreasing flux) \oviw, \osiivw, \nvw, \mgiiw\ and
\ciiw. Because \osiivw\ consists of a  blend of two emission doublets, 
it is unlikely that the corresponding absorption lines could be
disentangled. The other resonance lines left to consider are \ovi\
and \nv. In Fig.\,\ref{fig:osix} and \ref{fig:nfive}, we present the
column ratios \nohr\ and \nnhr, respectively, as a function of \up.
One can see from these figures that the ionization parameter could
be considerably better constrained if data on these resonance lines
were obtained. Thus, obtaining high-resolution optical spectra over
all emission lines is essential to better constrain the properties of 
these haloes.


\begin{figure}    
\resizebox{\hsize}{!}{\includegraphics{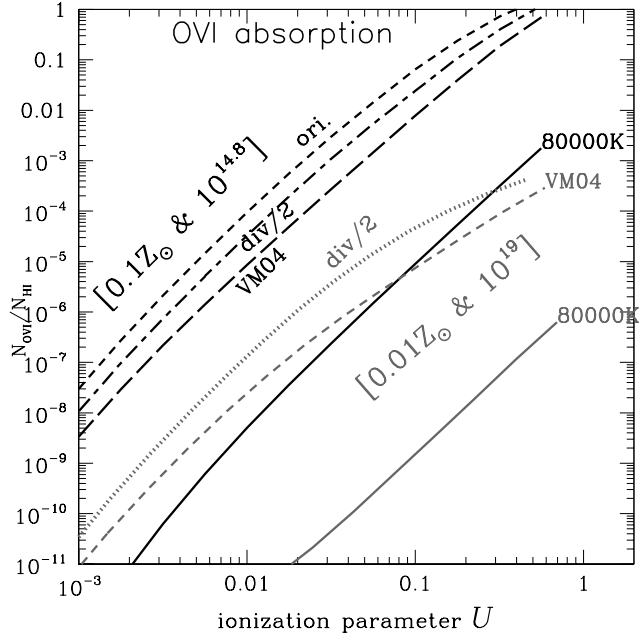}}
\caption[]{The column ratio \nohr\ derived from photoionization  by  four different  
\seds\ discussed  in Sect.\,\ref{sec:resu}, as a function of \up. }
\label{fig:osix}
\end{figure} 

\begin{figure}    
\resizebox{\hsize}{!}{\includegraphics{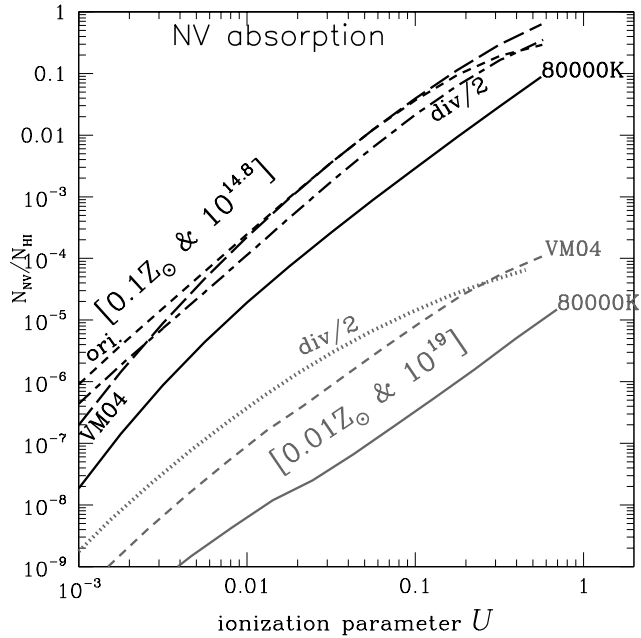}}
\caption[]{The column ratio \nnhr\ derived from photoionization  by  four different  
\seds\ discussed  in Sect.\,\ref{sec:resu}, as a function of \up. }
\label{fig:nfive}
\end{figure} 

\begin{acknowledgements}
One of the authors (LB) acknowledges financial support from CONACyT
grant 40096-E and the UNAM PAPIIT grants 113002 and 118601. RJW and
MJJ acknowlege the support of PPARC PDRAs. RAEF is affiliated to the
Research and Science Support Department of the European Space
Agency. Diethild Starkmeth helped us with proof-reading. We
acknowledge the technical support of Liliana Hern\'andez and Carmelo
Guzm\'an for configuring the Linux workstation Deneb.

\end{acknowledgements}


\end{document}